\documentclass[linenumbers]{aastex631}
\usepackage{amsmath}

\begin{document}
\nolinenumbers
\title{Search for joint multimessenger signals from potential galactic cosmic-ray accelerators with HAWC and IceCube }

\correspondingauthor{Kwok Lung Fan}
\email{klfan@umd.edu}

\correspondingauthor{IceCube Collaboration}
\email{analysis@icecube.wisc.edu}

\author{R.~Alfaro}
\affiliation{Instituto de F\'{i}sica, Universidad Nacional Autónoma de México, Ciudad de Mexico, Mexico }

\author{C.~Alvarez}
\affiliation{Universidad Autónoma de Chiapas, Tuxtla Gutiérrez, Chiapas, México}

\author{J.C.~Arteaga-Velázquez}
\affiliation{Universidad Michoacana de San Nicolás de Hidalgo, Morelia, Mexico }

\author{D.~Avila Rojas}
\affiliation{Instituto de F\'{i}sica, Universidad Nacional Autónoma de México, Ciudad de Mexico, Mexico }
\author[0000-0002-2084-5049]{H.A.~Ayala Solares}
\affiliation{Dept. of Physics, Pennsylvania State University, University Park, PA 16802, USA}

\author{R.~Babu}
\affiliation{Department of Physics and Astronomy, Michigan State University, East Lansing, MI, USA}

\author{E.~Belmont-Moreno}
\affiliation{Instituto de F\'{i}sica, Universidad Nacional Autónoma de México, Ciudad de Mexico, Mexico }
\author{K.S.~Caballero-Mora}
\affiliation{Universidad Autónoma de Chiapas, Tuxtla Gutiérrez, Chiapas, México}
\author[0000-0003-2158-2292]{T.~Capistrán}
\affiliation{Instituto de Astronom\'{i}a, Universidad Nacional Autónoma de México, Ciudad de Mexico, Mexico }

\author{A.~Carramiñana}
\affiliation{Instituto Nacional de Astrof\'{i}sica, Óptica y Electrónica, Puebla, Mexico }
\author[0000-0002-6144-9122]{S.~Casanova}
\affiliation{Institute of Nuclear Physics Polish Academy of Sciences, PL-31342 IFJ-PAN, Krakow, Poland }
\author[0000-0002-7607-9582]{U.~Cotti}
\affiliation{Universidad Michoacana de San Nicolás de Hidalgo, Morelia, Mexico }
\author[0000-0002-1132-871X]{J.~Cotzomi}
\affiliation{Facultad de Ciencias F\'{i}sico Matemáticas, Benemérita Universidad Autónoma de Puebla, Puebla, Mexico }
\author{S.~Coutiño de León}
\affiliation{Dept. of Physics and Wisconsin IceCube Particle Astrophysics Center, University of Wisconsin{\textemdash}Madison, Madison, WI 53706, USA}

\author[0000-0001-9643-4134]{E.~De la Fuente}
\affiliation{Departamento de F\'{i}sica, Centro Universitario de Ciencias Exactase Ingenierias, Universidad de Guadalajara, Guadalajara, Mexico }
\author{D.~Depaoli}
\affiliation{Max-Planck Institute for Nuclear Physics, 69117 Heidelberg, Germany}

\author{N.~Di Lalla}
\affiliation{Department of Physics, Stanford University: Stanford, CA 94305–4060, USA}

\author{R.~Diaz Hernandez}
\affiliation{Instituto Nacional de Astrof\'{i}sica, Óptica y Electrónica, Puebla, Mexico }

\author[0000-0002-0087-0693]{J.C.~Díaz-Vélez}
\affiliation{Dept. of Physics and Wisconsin IceCube Particle Astrophysics Center, University of Wisconsin{\textemdash}Madison, Madison, WI 53706, USA}
\author[0000-0001-5737-1820]{K.~Engel}
\affiliation{Department of Physics, University of Maryland, College Park, MD, USA}

\author{T.~Ergin}
\affiliation{Department of Physics and Astronomy, Michigan State University, East Lansing, MI, USA}

\author[0000-0002-8246-4751]{K.L.~Fan}
\affiliation{Department of Physics, University of Maryland, College Park, MD, USA}
\author[0000-0002-5387-8138]{K.~Fang}
\affiliation{Dept. of Physics and Wisconsin IceCube Particle Astrophysics Center, University of Wisconsin{\textemdash}Madison, Madison, WI 53706, USA}
\author{N.~Fraija}
\affiliation{Instituto de Astronom\'{i}a, Universidad Nacional Autónoma de México, Ciudad de Mexico, Mexico }

\author{S.~Fraija}
\affiliation{Instituto de Astronom\'{i}a, Universidad Nacional Autónoma de México, Ciudad de Mexico, Mexico }
\author[0000-0002-4188-5584]{J.A.~García-González}
\affiliation{Instituto Tecnológico y de Estudios Superiores de Monterrey - Campus Toluca: Toluca de Lerdo, Estado de México, MX}
\author[0000-0003-1122-4168]{F.~Garfias}
\affiliation{Instituto de Astronom\'{i}a, Universidad Nacional Autónoma de México, Ciudad de Mexico, Mexico }
\author[0000-0002-5209-5641]{M.M.~González}
\affiliation{Instituto de Astronom\'{i}a, Universidad Nacional Autónoma de México, Ciudad de Mexico, Mexico }
\author[0000-0002-9790-1299]{J.A.~Goodman}
\affiliation{Department of Physics, University of Maryland, College Park, MD, USA}
\author{S.~Groetsch}
\affiliation{Department of Physics, Michigan Technological University, Houghton, MI, USA }
\author{J.P.~Harding}
\affiliation{Physics Division, Los Alamos National Laboratory, Los Alamos, NM, USA }
\author{S.~Hernández-Cadena}
\affiliation{Tsung-Dao Lee Institute \& School of Physics and Astronomy, Shanghai Jiao Tong University, Shanghai, China}
\author{I.~Herzog}
\affiliation{Department of Physics and Astronomy, Michigan State University, East Lansing, MI, USA}

\author[0000-0002-5447-1786]{D.~Huang}
\affiliation{Department of Physics, University of Maryland, College Park, MD, USA}
\author[0000-0002-5527-7141]{F.~Hueyotl-Zahuantitla}
\affiliation{Universidad Autónoma de Chiapas, Tuxtla Gutiérrez, Chiapas, México}
\author{P.~Hüntemeyer}
\affiliation{Department of Physics, Michigan Technological University, Houghton, MI, USA }
\author[0000-0001-5811-5167]{A.~Iriarte}
\affiliation{Instituto de Astronom\'{i}a, Universidad Nacional Autónoma de México, Ciudad de Mexico, Mexico }
\author{S.~Kaufmann}
\affiliation{Universidad Politécnica de Pachuca, Pachuca, Hgo, Mexico }

\author[0000-0002-2467-5673]{J.~Lee}
\affiliation{Natural Science Research Institute, University of Seoul, Seoul, Republic of Korea}
\author[0000-0001-5516-4975]{H.~León Vargas}
\affiliation{Instituto de F\'{i}sica, Universidad Nacional Autónoma de México, Ciudad de Mexico, Mexico }

\affiliation{Instituto Nacional de Astrof\'{i}sica, Óptica y Electrónica, Puebla, Mexico }
\author[0000-0003-2810-4867]{G.~Luis-Raya}
\affiliation{Universidad Politécnica de Pachuca, Pachuca, Hgo, Mexico }

\author[0000-0001-8088-400X]{K.~Malone}
\affiliation{Physics Division, Los Alamos National Laboratory, Los Alamos, NM, USA }
\author{J.~Martínez-Castro}
\affiliation{Centro de Investigaci\'on en Computaci\'on, Instituto Polit\'ecnico Nacional, M\'exico City, M\'exico.}
\author[0000-0002-2610-863X]{J.A.~Matthews}
\affiliation{Dept of Physics and Astronomy, University of New Mexico, Albuquerque, NM, USA }
\author{P.~Miranda-Romagnoli}
\affiliation{Universidad Autónoma del Estado de Hidalgo, Pachuca, Mexico }
\author{J.A.~Montes}
\affiliation{Instituto de Astronom\'{i}a, Universidad Nacional Autónoma de México, Ciudad de Mexico, Mexico }

\author{E.~Moreno}
\affiliation{Facultad de Ciencias F\'{i}sico Matemáticas, Benemérita Universidad Autónoma de Puebla, Puebla, Mexico }

\author[0000-0002-7675-4656]{M.~Mostafá}
\affiliation{Department of Physics, Temple University, Philadelphia, Pennsylvania, USA}
\author[0000-0003-1059-8731]{L.~Nellen}
\affiliation{Instituto de Ciencias Nucleares, Universidad Nacional Autónoma de Mexico, Ciudad de Mexico, Mexico }
\author[0000-0002-6859-3944]{M.U.~Nisa}
\affiliation{Department of Physics and Astronomy, Michigan State University, East Lansing, MI, USA}
\author[0000-0002-5448-7577]{N.~Omodei}
\affiliation{Department of Physics, Stanford University: Stanford, CA 94305–4060, USA}

\author{M.~Osorio}
\affiliation{Instituto de Astronom\'{i}a, Universidad Nacional Autónoma de México, Ciudad de Mexico, Mexico }

\author[0000-0002-8774-8147]{Y.~Pérez Araujo}
\affiliation{Instituto de Astronom\'{i}a, Universidad Nacional Autónoma de México, Ciudad de Mexico, Mexico }
\author[0000-0001-5998-4938]{E.G.~Pérez-Pérez}
\affiliation{Universidad Politécnica de Pachuca, Pachuca, Hgo, Mexico }
\author[0000-0002-6524-9769]{C.D.~Rho}
\affiliation{Department of Physics, Sungkyunkwan University, Suwon 16419, South Korea}
\author[0000-0003-1327-0838]{D.~Rosa-González}
\affiliation{Instituto Nacional de Astrof\'{i}sica, Óptica y Electrónica, Puebla, Mexico }

\author{H.~Salazar}
\affiliation{Facultad de Ciencias F\'{i}sico Matemáticas, Benemérita Universidad Autónoma de Puebla, Puebla, Mexico }

\author{D.~Salazar-Gallegos}
\affiliation{Department of Physics and Astronomy, Michigan State University, East Lansing, MI, USA}
\author{A.~Sandoval}
\affiliation{Instituto de F\'{i}sica, Universidad Nacional Autónoma de México, Ciudad de Mexico, Mexico }
\author{M.~Schneider}
\affiliation{Department of Physics, University of Maryland, College Park, MD, USA}
\author{J.~Serna-Franco}
\affiliation{Instituto de F\'{i}sica, Universidad Nacional Autónoma de México, Ciudad de Mexico, Mexico }
\author{A.J.~Smith}
\affiliation{Department of Physics, University of Maryland, College Park, MD, USA}
\author{Y.~Son}
\affiliation{Natural Science Research Institute, University of Seoul, Seoul, Republic of Korea}

\author{O.~Tibolla}
\affiliation{Universidad Politécnica de Pachuca, Pachuca, Hgo, Mexico }
\author[0000-0001-9725-1479]{K.~Tollefson}
\affiliation{Department of Physics and Astronomy, Michigan State University, East Lansing, MI, USA}
\author[0000-0002-1689-3945]{I.~Torres}
\affiliation{Instituto Nacional de Astrof\'{i}sica, Óptica y Electrónica, Puebla, Mexico }
\author{R.~Torres-Escobedo}
\affiliation{Tsung-Dao Lee Institute \& School of Physics and Astronomy, Shanghai Jiao Tong University, Shanghai, China}
\author{R.~Turner}
\affiliation{Department of Physics, Michigan Technological University, Houghton, MI, USA }

\author{F.~Ureña-Mena}
\affiliation{Instituto Nacional de Astrof\'{i}sica, Óptica y Electrónica, Puebla, Mexico }

\author{X.~Wang}
\affiliation{Department of Physics, Michigan Technological University, Houghton, MI, USA }

\author{I.J.~Watson}
\affiliation{Natural Science Research Institute, University of Seoul, Seoul, Republic of Korea}

\author{K.~Whitaker}
\affiliation{Dept. of Physics, Pennsylvania State University, University Park, PA 16802, USA}

\author{E.~Willox}
\affiliation{Department of Physics, University of Maryland, College Park, MD, USA}

\author{H.~Wu}
\affiliation{Dept. of Physics and Wisconsin IceCube Particle Astrophysics Center, University of Wisconsin{\textemdash}Madison, Madison, WI 53706, USA}
\author[0009-0006-3520-3993]{S.~Yu}
\affiliation{Dept. of Physics, Pennsylvania State University, University Park, PA 16802, USA}
\author[0000-0002-9307-0133]{S.~Yun-Cárcamo}
\affiliation{Department of Physics, University of Maryland, College Park, MD, USA}
\author{H.~Zhou}
\affiliation{Tsung-Dao Lee Institute \& School of Physics and Astronomy, Shanghai Jiao Tong University, Shanghai, China}
\author{C.~de León}
\affiliation{Universidad Michoacana de San Nicolás de Hidalgo, Morelia, Mexico }

\collaboration{418}{HAWC Collaboration}

\author[0000-0001-6141-4205]{R. Abbasi}
\affiliation{Department of Physics, Loyola University Chicago, Chicago, IL 60660, USA}

\author[0000-0001-8952-588X]{M. Ackermann}
\affiliation{Deutsches Elektronen-Synchrotron DESY, Platanenallee 6, D-15738 Zeuthen, Germany}

\author{J. Adams}
\affiliation{Dept. of Physics and Astronomy, University of Canterbury, Private Bag 4800, Christchurch, New Zealand}

\author[0000-0002-9714-8866]{S. K. Agarwalla}
\altaffiliation{also at Institute of Physics, Sachivalaya Marg, Sainik School Post, Bhubaneswar 751005, India}
\affiliation{Dept. of Physics and Wisconsin IceCube Particle Astrophysics Center, University of Wisconsin{\textemdash}Madison, Madison, WI 53706, USA}

\author[0000-0003-2252-9514]{J. A. Aguilar}
\affiliation{Universit{\'e} Libre de Bruxelles, Science Faculty CP230, B-1050 Brussels, Belgium}

\author[0000-0003-0709-5631]{M. Ahlers}
\affiliation{Niels Bohr Institute, University of Copenhagen, DK-2100 Copenhagen, Denmark}

\author[0000-0002-9534-9189]{J.M. Alameddine}
\affiliation{Dept. of Physics, TU Dortmund University, D-44221 Dortmund, Germany}

\author{N. M. Amin}
\affiliation{Bartol Research Institute and Dept. of Physics and Astronomy, University of Delaware, Newark, DE 19716, USA}

\author[0000-0001-9394-0007]{K. Andeen}
\affiliation{Department of Physics, Marquette University, Milwaukee, WI 53201, USA}

\author[0000-0003-4186-4182]{C. Arg{\"u}elles}
\affiliation{Department of Physics and Laboratory for Particle Physics and Cosmology, Harvard University, Cambridge, MA 02138, USA}

% ...

\collaboration{418}{IceCube Collaboration}


%% encouraged to surround collaboration identifiers with~()s. The 

\begin{abstract}
The origin of high-energy galactic cosmic rays is yet to be understood, but some galactic cosmic-ray accelerators can accelerate cosmic rays up to PeV energies. The high-energy cosmic rays are expected to interact with the surrounding material or radiation, resulting in the production of gamma rays and neutrinos. To optimize for the detection of such associated production of gamma rays and neutrinos for a given source morphology and spectrum,  a multi-messenger analysis that combines gamma rays and neutrinos is required. In this study, we use the Multi-Mission Maximum Likelihood framework~(3ML) with IceCube Maximum Likelihood Analysis software~(i3mla) and HAWC Accelerated Likelihood~(HAL) to search for a correlation between 22 known gamma-ray sources from the third HAWC gamma-ray catalog and 14 years of IceCube track-like data. No significant neutrino emission from the direction of the HAWC sources was found. We report the best-fit gamma-ray model and 90\% CL neutrino flux limit from the 22 sources. From the neutrino flux limit, we conclude that for five of the sources, the gamma-ray emission observed by HAWC cannot be produced purely from hadronic interactions. We report the limit for the fraction of gamma rays produced by hadronic interactions for these five sources.

\end{abstract}



\section{Introduction} \label{sec:intro}
Cosmic rays~(CRs) were discovered more than 100 years ago \citep{Hess:1912srp},  but where these high-energy particles come from has not yet been well understood. The highest energy charged particles, protons and nuclei, are thought to be extraterrestrial in origin, and their spectrum can be parametrized using a power law with an index of 2.7 up to ${\rm E}_{\rm CR}\sim 3$~PeV (or the ``knee'' in the spectrum), above which the spectrum steepens \citep{cr_knee_1984,cr_sp,cr_lhaaso,CR_tibet,cr_icetop,cr_spectrum_casamia}. The CRs below the knee are thought to originate from sources within the Milky Way galaxy, hinting at the presence of PeV hadron accelerators within the galaxy known as Galactic PeVatrons \citep{first_pevatron}. Galactic PeVatrons are expected to accelerate particles up to PeV energies through Fermi acceleration and these accelerated particles are expected to be confined within the Milky Way galaxy \citep{Pevatron_theory}. The AS gamma collaboration reported the detection of diffuse gamma rays from the galactic disk with energy between 100 TeV and 1 PeV, hinting at the presence of PeVatrons within our galaxy \citep{as_gamma}.

Supernova remnants~(SNRs) are a potential class of Galactic PeVatrons. The particles are accelerated by diffusive shock acceleration in the expanding shock. Only 10\% of the energy of SNRs is needed to account for the observed CR energy density in the Galaxy \citep{Blandford}. For the most energetic galactic CRs, it is assumed that only young SNRs can accelerate CRs beyond 100 TeV \citep{youngsnr_aspev}. The accelerated CRs can escape the SNR and diffuse into the surrounding interstellar medium. They can interact with the surrounding material like molecular clouds and produce charged and neutral pions. Therefore, young SNRs located near star clusters or molecular clouds are thought to be bright in gamma rays due to hadronic interactions.  High-energy gamma rays and neutrinos can be efficiently produced by the decay of charged and neutral pions. Other potential classes of Galactic PeVatrons include young massive star clusters, YMCs \citep{ymc_nu} and pulsar wind nebulae, PWNe \citep{pwn_nu}. 

TeV gamma rays can also be produced by leptonic interactions such as inverse Compton scattering. However, inverse Compton scattering is strongly suppressed at energies above $50$~TeV \citep{TeVastro} due to the Klein-Nishina effect \citep{KN_effect}. On the other hand, the gamma rays produced by neutral pion decays are not suppressed at O(10 TeV), making TeV gamma-ray emitters excellent candidates for Galactic PeVatrons. Moreover, high-energy neutrinos are also produced in hadronic interactions via charged pion decays and their detection is evidence of hadron acceleration. Because of this relationship, a joint search using TeV gamma-ray and neutrino data from known TeV gamma-ray emitters can test whether a source is a Galactic PeVatron.
% On the other hand, leptonic processes like inverse Compton scattering can also produce high-energy gamma rays. However, inverse Compton scattering is heavily suppressed at energy higher than $50$~TeV \citep{TeVastro} due to Klein-Nishina suppression and is inefficient in producing high-energy gamma rays. Therefore, TeV gamma-ray emitters are excellent candidates for Galactic PeVatrons and neutrino signals will be the smoking gun of hadronic acceleration. Motivated by this relation, a joint search using TeV gamma-ray and neutrino data on known TeV gamma-ray emitters can test the hypothesis of whether a source is Galactic PeVatrons.

Several potential Galactic PeVatrons have been investigated by the community. The HAWC collaboration reported that the TeV gamma-ray spectrum of HAWC J1825-134 extends up to 200 TeV without any cut-off, much higher than the Klein-Nishina suppression energy \citep{j1825}. This may indicate
 hadronic acceleration. The HAWC collaboration also reported HAWC J2227+610 as a potential PeVatron candidate from TeV gamma-ray observations and concluded the proton cutoff energy is at least 800 TeV \citep{j2227}. The HAWC collaboration studied the ultra-high-energy spectrum of MGRO J1908+06 and their model allows a hadronic component at the highest energy \citep{1908}. The HESS Collaboration reported evidence of a PeVatron at the galactic center based on gamma-ray observations up to 40 TeV \citep{HESS2016}. The LHAASO collaboration reported an ultrahigh-energy gamma ray bubble up to a few PeV from the direction of the Cygnus X region, and suggested that a super PeVatron with maximum proton energy of 10 PeV could power the source \citep{Cygnus}.  The LHAASO collaboration detected 12 gamma-ray sources with emissions above 100 TeV \citep{cao2021ultrahigh} and recently reported their first catalog 1LHAASO \citep{LHAASO_first_catalog}. The 1LHAASO contains 90 gamma-ray sources, and 43 sources have emission above 100 TeV, making them possible Galactic PeVatrons.  %The detection of astrophysical neutrino for these potential Galactic PeVatrons will be the smoking gun of hadronic acceleration.

Many efforts have been made to search for PeVatrons using neutrino data. \citet{Huang_2022} set a constraint on neutrino emission from LHAASO's first catalog of sources with energy above 100 TeV \citep{cao2021ultrahigh} using previous IceCube neutrino upper limits. They used measurements from other gamma-ray instruments and constrained the neutrino flux using the measured gamma-ray spectrum. Another search by \citet{huang_li_mnras} used 10 years of public
IceCube track data to search for neutrino emission from the same LHAASO catalog \citep{cao2021ultrahigh} with fixed morphology and different spectral models. The IceCube Neutrino Observatory has previously searched for neutrino emission from galactic sources, both individual sources and classes of sources. \citet{Fan_chang_icecube} searched for neutrino emission from LHAASO's first catalog of sources with energy above 100 TeV \citep{cao2021ultrahigh} using a fixed spatial extension and a power law spectrum and 12 years of IceCube track data. They found no significant emission, neither from single sources nor from stacked catalogs of SNRs and PWNe. Recently, IceCube has detected neutrinos from the galactic plane \citep{icecube_gp}. The observation was made using a neutrino detection channel, cascades, which has relatively large angular uncertainty. Therefore, IceCube could not distinguish between diffuse emission in the galactic plane and a collection of individual sources. The IceCube collaboration also searched the entire galactic plane and regions near TeV gamma-ray sources using an extended source hypothesis and a power law spectrum, finding no significant neutrino emission \citep{icecube_extend}. However, these searches have not directly incorporated gamma-ray data. The simultaneous use of both data sets incorporates more physically motivated spectra and more accurate morphology. In this work, for the first time, we simultaneously use both gamma-ray and neutrino data to improve the sensitivity of the search and provide a direct constraint on the hadronic ratio, i.e., the fraction of gamma rays originating from hadronic interactions. The improvement in sensitivity of the search comes from the additional 2 years for data taken with full detector compared to \citet{Fan_chang_icecube} and the modeling of the neutrino emission under the assumption that the gamma rays and neutrinos originated from the same population of hadrons. 

%However, none of these searches has incorporated the gamma-ray data from other experiments to perform a more physically motivated search and provide a neutrino flux limit based on the gamma-ray observation.

If the gamma rays and neutrinos originate from hadronic interactions, the spectra of the gamma rays and neutrinos have a fixed relationship. In the case of Galactic PeVatrons, the proton-proton~(pp) interaction is expected to be dominant as compared to the photo-hadronic interactions \citep{nu_gamma_relation}. In the pp interactions, protons interact with nearby material and produce charged and neutral pions. The ratio between the production of charged and neutral pions is about 2:1. Charged pions decay further into three neutrinos and other products, with each resulting neutrino having about a quarter of the energy of the charged pion. The neutral pions decay into two gamma rays, each with half of the energy of the neutral pion, on average. Therefore, we expect that for charged and neutral pion decay, the energy of the neutrino produced will be half the energy of the gamma-ray produced on average. On Earth, the observed flux is
\begin{equation}
E_\gamma J_\gamma(E_\gamma) \approx e^{\frac{-d}{\lambda_{\gamma\gamma}}} \frac{1}{3}\sum_{\nu_\alpha} E_\nu J_{\nu_\alpha}(E_\nu)
\end{equation}
where $J_{\nu_\alpha}(E_\nu)$ is the neutrino differential particle flux with the neutrino flavor $\alpha$, $d$ is the distance to the source, $\lambda_{\gamma\gamma}$ is the interaction length of the gamma ray and $J_\gamma(E_\gamma)$ is the gamma-ray differential particle flux \citep{nu_gamma_relation}. Since we only consider galactic sources, we assume that the absorption of gamma-rays is negligible~($\frac{-d}{\lambda_{\gamma\gamma}} \approx 0$) and the flux from each neutrino flavor is the same due to oscillations. The relationship between the gamma-ray flux ($J_\gamma$) and the muon neutrino flux ($J_{\nu_\mu}$) is thus
\begin{equation}
\label{flux_convert}
J_{\nu_\mu}(E_{\nu_\mu}) = 2 J_\gamma(2E_{\nu_\mu}).
\end{equation}
Therefore, if the gamma rays from a galactic source originate from hadronic interactions, the neutrino counterpart should exhibit the same morphology and have the same spectral shape shifted to lower energies. This relationship is used in this work to model and simultaneously fit the gamma-ray and neutrino data.

%\hyperref[sec:instruments]{Section~\ref*{sec:instruments}} introduces the HAWC Gamma-ray Observatory, IceCube Neutrino Observatory, and the datasets used in this study. \hyperref[sec:analysis]{Section~\ref*{sec:analysis}} describes the method for HAWC source analysis with HAWC data and the joint analysis using both HAWC and IceCube data. \hyperref[sec:result]{Section~\ref*{sec:result}} shows the result of the gamma rays fit using HAWC data and neutrino flux limit obtained from the joint analysis. Finally, \hyperref[sec:result]{section~\ref*{sec:result}} summarizes the findings and significance of this study.

\section{Instruments and Datasets}
\label{sec:instruments}
\subsection{HAWC Gamma-ray Observatory}
The High Altitude Water Cherenkov~(HAWC) Observatory is located near Sierra Negra, Mexico, at a latitude of $19^\circ N $ and an altitude of 4100 m. It has been in continuous operation since November 2014 \citep{HAWC_NIM}. The observatory consists of 300 Water Cherenkov Detectors~(WCDs), covering an area of 22000 $\text{m}^2$. Each WCD consists of a steel
tank 7.3 m in diameter and 5.4 m high,  lined with a plastic bladder and filled with purified water. At the bottom of the water tank, three 8-inch photomultiplier tubes~(PMTs) are arranged in an equilateral triangle with a side length of 3.2m, and with one 10-inch PMT in the center \citep{HAWC_NIM}.

When high-energy gamma rays interact with the Earth's atmosphere, they create a cascade of particles called an extensive air shower~(EAS). When these particles reach the HAWC detector and pass through the water at a speed greater than the speed of light in water, Cherenkov light is induced and collected by the PMTs in the WCDs. HAWC uses the observed Cherenkov emission to reconstruct the properties of the primary gamma rays. 

HAWC divides the detected shower events into a 2D array of analysis bins based on the fraction of triggered PMTs during a shower event and the reconstructed energy of the event \citep{crab_2019}. The events within an analysis bin are expected to have similar characteristics, including angular resolution, primary gamma-ray energy, and gamma-to-hadron likelihood~(signal-to-noise ratio), which allows HAWC to perform a binned likelihood analysis. 

In this study, we use 2141 days of HAWC data using the Pass 5 reconstruction \citep{pass5paper}. The event-related energy is estimated using the ``ground parameter" estimator described in \cite{crab_2019}. This energy estimator uses the charge at a fixed optimal distance from the shower axis to estimate the energy. The dataset covers the period from June 15, 2015 to October 21, 2021 with a livetime of more than 90\%. The dataset contains 29\% more data with an improved reconstruction compared to the dataset used in the third HAWC catalog (3HWC) \citep{3hwc}.

\subsection{The IceCube Neutrino Observatory}
The IceCube Neutrino Observatory is a neutrino detector located at the geographic South Pole \citep{IceCube_ins}. The primary IceCube detector comprises a cubic kilometer array of 5160 digital optical modules~(DOMs). These DOMs are arranged along 86 readout and support cables, called strings, embedded in the glacial ice at depths between 1.45 km and 2.45 km. Each DOM consists of a 10-inch PMT and its readout electronics and is housed in a pressure-resistant sphere \citep{icecube_instrumentation}. DOMs are designed to detect the Cherenkov light produced by charged particles moving through the ice \citep{icecube_pmt}. These particles originate from the charged-current~(CC) interactions or neutral-current~(NC) interactions of all flavors of neutrinos. IceCube events are typically divided into two categories: track-like events produced by CC interactions of muon neutrinos and cascade-like events from all other CC interactions and NC interactions. The DOMs record the Cherenkov light, allowing events to be classified and event properties, such as the direction of the incident neutrinos and the deposited energy, to be reconstructed \citep{icecube_energy_reco}. %Apart from the main detector, IceCube also includes DeepCore, which is a GeV-scale neutrino sub-detector \citep{deepcore}, and IceTop, an air shower array primarily for cosmic rays detection \citep{icetop}. Both of which are not used in this study.

The dataset used in this analysis includes 14 years of IceCube track-like data from April 6, 2008 to May 23, 2022. Track-like data provide a better angular resolution~( $\sim 1^{\circ}$ angular uncertainty for a typical $\sim$ TeV neutrino) compared to cascade-like events. This makes track-like events more suited to search for sources with an extension below a few degrees. The data from April 6, 2008 to May 13, 2011 were taken with the partially constructed detector with 40, 59, and 79 strings~(IC40, IC59, and IC79), with each configuration spanning approximately one year \citep{icecube_time_2014}. The remaining data were recorded with the full detector configuration with 86 strings~(IC86).

\section{Analysis method}
\label{sec:analysis}
\subsection{Source selection}
The third HAWC catalog (3HWC) includes 65 sources with a significance above 5$\sigma$ based on 1523 days of data \citep{3hwc}. The 3HWC provides the location and the gamma-ray flux at a pivot energy of 7 TeV for each source in the catalog. The pivot energy $E_0$ is the energy of normalization of the spectral function, which is specified in terms of ${E}/{E_0}$. If the gamma-ray emission from these sources comes from hadronic interactions, the neutrino flux will be described by the same spectral function with half the pivot energy (3.5 TeV), according to Equation \ref{flux_convert}. We calculate the IceCube 90\% confidence level (CL) sensitivity for a point source with an $E^{-2}$ unbroken power law spectrum. The sensitivity is defined as the median expected 90\% CL upper limit. The sensitivity can be determined using background trials and signal trials. Background trials are done by performing the analysis using simulated data obtained by scrambling real data in right ascension~(RA). Scrambling data as background does not suffer from systematic uncertainties and therefore is a robust way to generate background trials. Signal trials are generated by scrambling data and adding signal using simulation data. The calculation of sensitivity is based on a maximum likelihood method as used in the IceCube 10-year search for neutrino sources\citep{10yearsIceCube}. In this work, we use a new likelihood analysis software component called i3mla \citep{i3mla} with the multi-mission maximum likelihood framework~(3ML) \citep{3ML}. Details of i3mla are described in section \ref{joint-fit}. The likelihood consists of a spatial term and an energy term, as explained in more detail in Appendix \ref{likelihood_appendix}.

\begin{figure}[h!]

\centering
\includegraphics[width=0.7\textwidth]{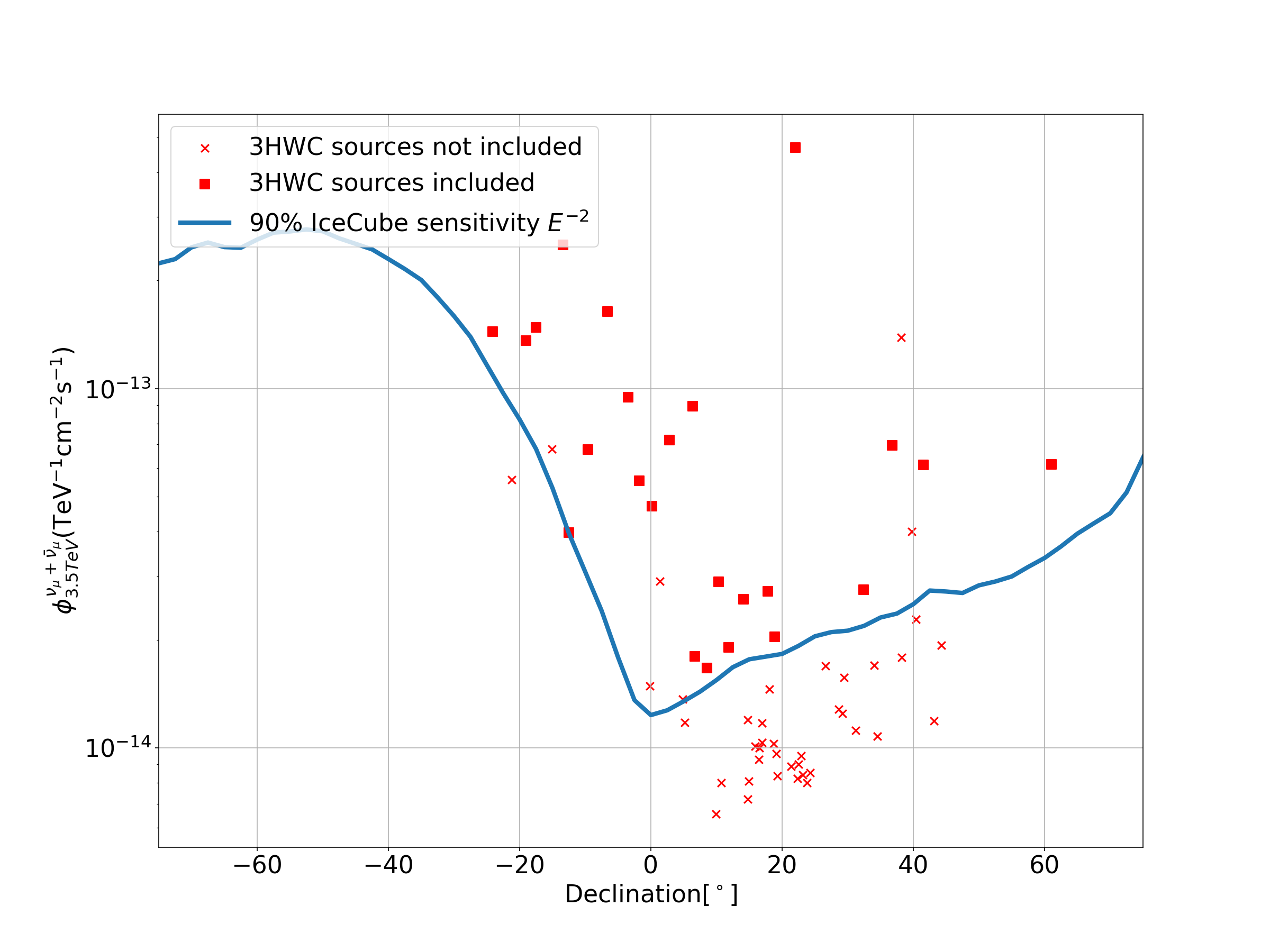}
\caption{The IceCube $E^{-2}$ time-integrated point source sensitivity as a function of source declination. The x-axis is the declination in degrees and the y-axis is the differential neutrino flux~($\mathrm{Te}V^{-1}\mathrm{cm}^{-2}\mathrm{s}^{-1}$) at pivot energy $E_0$ = 3.5 TeV. Each red cross and red square represents the predicted neutrino flux of a 3HWC source assuming all the gamma rays from the sources originated from proton-proton hadronic interaction. Galactic 3HWC sources (excluding 3HWC secondary sources) with a predicted neutrino flux above the IceCube sensitivity are included in our source list and shown as red squares. } %The blue line is the IceCube sensitivity assuming the neutrino source spectrum follows a $E^{-2}$ spectrum. }
\label{source_selection_plot}
\end{figure}

We select sources whose predicted neutrino flux from hadronic interactions is larger than the sensitivity of the IceCube $E^{-2}$ point source search. The source selection step is motivated by the trade-off between searching as many possible sources as possible to avoid missing potential sources and avoiding suffering sensitivity loss from the trial factor. The hard $E^{-2}$ spectrum is an optimistic choice made to avoid missing any potential sources in our source selection step since it is close to the hardest spectrum expected. The sensitivity of the joint search will depend on the best-fit gamma-ray spectrum and spatial extension and is different from the $E^{-2}$ point source sensitivity here. The sensitivity of IceCube and the predicted neutrino flux from 3HWC sources are shown in Figure \ref{source_selection_plot}. We exclude two extra-galactic sources, Markarian 421 and Markarian 501, as this study focuses on galactic sources. We also exclude four sources classified as secondary sources in 3HWC, as these are likely to be statistical fluctuations near bright sources. Our final source list comprises 22 sources from 3HWC. Many of the sources are still unidentified but some have potential associations with pulsar wind nebulae, supernova remnants, and star-forming regions.
%%% ref result table.

\subsection{Model selection and fit using gamma-ray data}
To obtain a spatial and spectral model for the joint analysis of the sources, we first fit each gamma-ray source with an updated HAWC data set. We use 2141 days of HAWC ground parameter energy estimator maps with the Pass 5 reconstruction for our model selection procedure. The dataset contains 29\% more data with an improved reconstruction compared to the dataset used in 3HWC. The new reconstruction provides a better angular reconstruction and gamma/hadron separation. We use the 3ML framework \citep{3ML} with the HAWC Accelerated Likelihood~(HAL) plugin \citep{HAL} for our model selection and model fitting. HAL is a Python-based, maximum likelihood software module designed to perform HAWC analysis using a binned likelihood. In conjunction with 3ML, it can perform both stand-alone HAWC analyses and joint analyses with other experiments. We select the model that best describes the gamma-ray emission based on the log-likelihood~(LLH). We use the position in 3HWC as a starting point for our initial fit. First, we use a simple power law and a point source as spectral and spatial hypotheses. The simple power law is parameterized as $\frac{dN_\gamma}{dE_\gamma} = \phi (\frac{E_\gamma}{7 {\rm TeV}})^{\alpha}$, where $\phi$ is the flux normalization and $\alpha$ is the spectral index. The free parameters in this model are the Right Ascension~(RA), the Declination~(Dec), the spectral index, and the gamma-ray flux. Second, we test the extended source hypothesis, where we fix the RA and Dec of the source to the best-fit values and fit for a Gaussian extension along with the spectral index and flux. We accept the extended source hypothesis if twice the difference between the log-likelihood ($2 \Delta LLH$) of the two models (point source hypothesis and extended source hypothesis) is greater than 16, corresponding to $4 \sigma$ according to Wilks' theorem \citep{wilks}. Finally, we test the spectrum curvature hypothesis, where we fixed the RA, Dec, and extension~(if the source is extended) and we use a log-parabola spectrum to fit the data again. The log-parabola spectrum is parameterized by 
\begin{equation}
\frac{dN_\gamma}{dE_\gamma} = \phi \cdot(\frac{E_\gamma}{7 {\rm TeV}})^{{\alpha}-\beta\cdot \log\frac{E_\gamma}{ {7\rm TeV}}}
\end{equation}
The free parameters in this model are $\alpha$, $\beta$ and the gamma-ray flux at 7 TeV. We accept the spectral curvature hypothesis if the $2 \Delta LLH$ between the spectral curvature hypothesis and the power law hypothesis is greater than 16. Based on the results of these fits, we obtain the final model that contains information about the morphology and spectrum of the source.
%%%ref result table

\subsection{IceCube likelihood with i3mla}
Combining IceCube neutrino data with the HAWC gamma-ray data is a non-trivial task because of the complexity of the instrument response functions and the likelihood calculation. We use a Python-based maximum likelihood module called i3mla \citep{i3mla} for IceCube likelihood calculation. Similar to HAL, i3mla is specifically designed to be compatible with the 3ML framework. i3mla uses instrument response functions~(IRFs) to encapsulate the detector properties and performances of the detector determined from Monte Carlo simulations. The IRFs then can be used to calculate the energy likelihood of IceCube. This differs from other IceCube analysis software that pre-computes the energy likelihood for an assumed spectrum, e.g. a simple power law, and uses all the Monte Carlo data for the energy likelihood calculation. The new tabulated IRFs method speeds up the likelihood computation process when fitting for an arbitrary spectrum, making it possible to fit any spectral parameters during the model fitting in i3mla.

In this analysis, we use spatial and energy likelihood terms to calculate the likelihood of the model. The signal hypothesis assumes some neutrino emission from the source of interest. The background hypothesis assumes all non-neutrino events and neutrino events originate from atmospheric muons, atmospheric neutrinos, and astrophysical neutrinos that are not associated with the source. We model the IceCube point spread function~(PSF) as a 2D Gaussian and the angular error of the 2D Gaussian is estimated during the reconstruction. The signal spatial likelihood is the PSF convolved with the morphology of the source and the signal energy likelihood is calculated using the IRFs and the model spectrum. The background spatial likelihood is estimated from the data by binning the data in reconstructed declination and creating a spline using the resulting histogram. The background spatial likelihood is a function of declination only. The background energy likelihood is estimated from the data by weighting the simulation using the spectral hypothesis and binning it in reconstructed declination and energy. The method is conceptually identical to the previous IceCube 10-year time-integrated point source search in \citet{10yearsIceCube} and the extended source search in \citet{Fan_chang_icecube,icecube_extend}. A detailed description of the likelihood can be found in Appendix \ref{likelihood_appendix}. 
\subsection{Joint fit with Gamma rays and Neutrinos}
\label{joint-fit}
After obtaining a spatial and spectral model for each source, we perform a joint fit with HAWC gamma-ray data and IceCube neutrino data. We keep the location and extension of the model fixed. We add a neutrino component to the model that coincides with the gamma-ray source. We connect the neutrino and gamma-ray emission spectral parameters according to the relation in Equation \ref{flux_convert} and allow them to float during the fit. We let the flux normalization of the neutrino and gamma-ray components float independently. Table \ref{relation} shows the properties of the model parameters.

\begin{table}[!ht]
    
    \centering
    \begin{tabular}{cccc}
    \hline
        Model parameters & Gamma-ray spectrum & Neutrino spectrum & Relation \\ \hline
        Right ascension(RA) & Fix & Fix & Equal \\ \hline
        Declination(dec) & Fix & Fix & Equal \\ \hline
        Extension & Fix & Fix & Equal \\ \hline
        $\alpha$ & Float & Float & Equal \\ \hline
        $\beta$ & Float & Float & Equal \\ \hline
        Pivot energy & Fix & Fix & Neutrino pivot energy = Gamma-ray pivot energy / 2 \\ \hline
        Flux normalization & Float & Float & Independent \\ \hline
    \end{tabular}
    \caption{Table summarizing the model parameters of the Gamma-ray and Neutrino components and their relations. Parameters listed as fix will be fixed during model fitting and parameters listed as float will be fitted during model fitting.}  
    \label{relation}
\end{table}

%, rather than using the entire Monte Carlo dataset, thereby accelerating the likelihood computation process. This allows the program to compute the energy likelihood during the model-fitting process instead of pre-computing it.

To calculate the p-value, we perform a large number of background trials by scrambling the IceCube data in RA and building the background test statistic~(TS) distribution \citep{icecube_time_integrated}. This is done to avoid any dependence on the simulation when testing for a neutrino source. The calculation of the IceCube component of the likelihood is fast. However, using raw HAWC binned maps for computing the HAWC component of the likelihood is prohibitively computationally expensive for such a large number of trials. Therefore, the HAWC data are fixed during the background scrambling process so that we can pre-compute the HAWC component of the likelihood. We extract the HAWC likelihood around the best-fit spectral parameters and save it into a table. Since the location and extension are fixed during the fit, we do not extract the likelihood around the spatial parameters. For spectral parameters, we evaluate the HAWC likelihood within $\pm 5\sigma$ of the best-fit spectral shape parameters while maximizing it over the flux normalization.

\begin{figure}[!ht]

\centering
\includegraphics[width=0.6\textwidth]{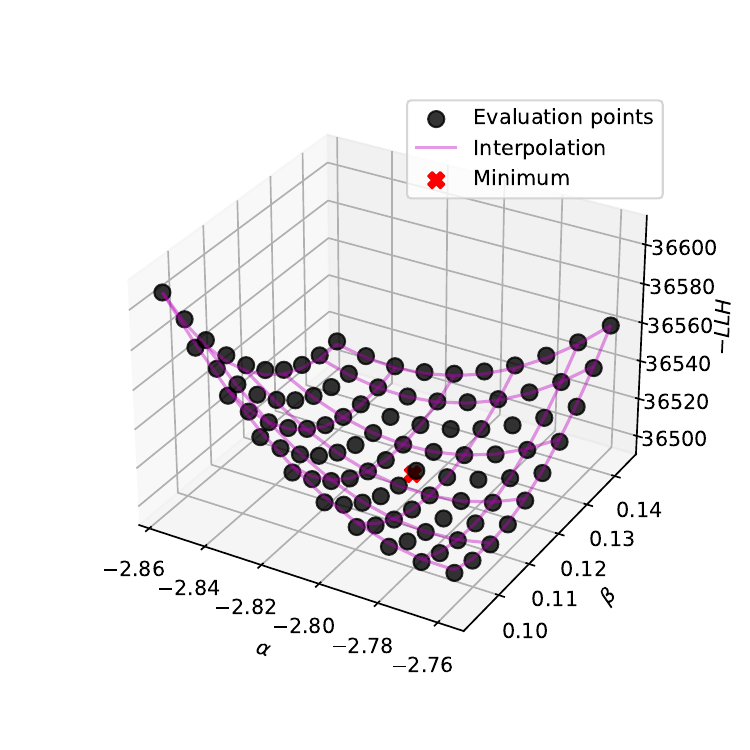}
\caption{A plot visualizing the HAWC likelihood for an example source. The y axis is the negative LLH. The negative LLH is evaluated around the best-fit model. For a log-parabola spectrum, we evaluate 50 points within $\pm 5 \sigma$ for $\alpha$ and $\beta$ around the minimum. For a power-law spectrum, we evaluate 300 points for spectral index within $\pm 5 \sigma$. The negative log-likelihood was minimized over the gamma-ray flux for each point.}
\label{llhplot}
\end{figure}

Since the gamma-ray and neutrino fluxes are floated independently, maximizing the likelihood over the gamma-ray flux normalization leads to a maximum likelihood for HAWC at the spectral shape parameters in the joint fit. For a log-parabola spectrum, there are two free parameters for the spectral shape~($\alpha$ and $\beta$), and we use a 50$\times$50 grid around the best-fit parameters for the likelihood evaluation. For a power-law spectrum, there is only one free parameter for the spectral shape~($\alpha$) and we evaluate 300 points around the best-fit parameters. We have developed a new custom plugin for 3ML to read the likelihood table and construct a spline. The 3ML plugin then returns the spline values for the
likelihood evaluations. This method speeds up the likelihood calculation because the program does not have to calculate the HAWC likelihood for each pixel and each bin during the fitting. Figure \ref{llhplot} shows an example of the likelihood.

\begin{figure}[!ht]

\centering
\includegraphics[width=\textwidth]{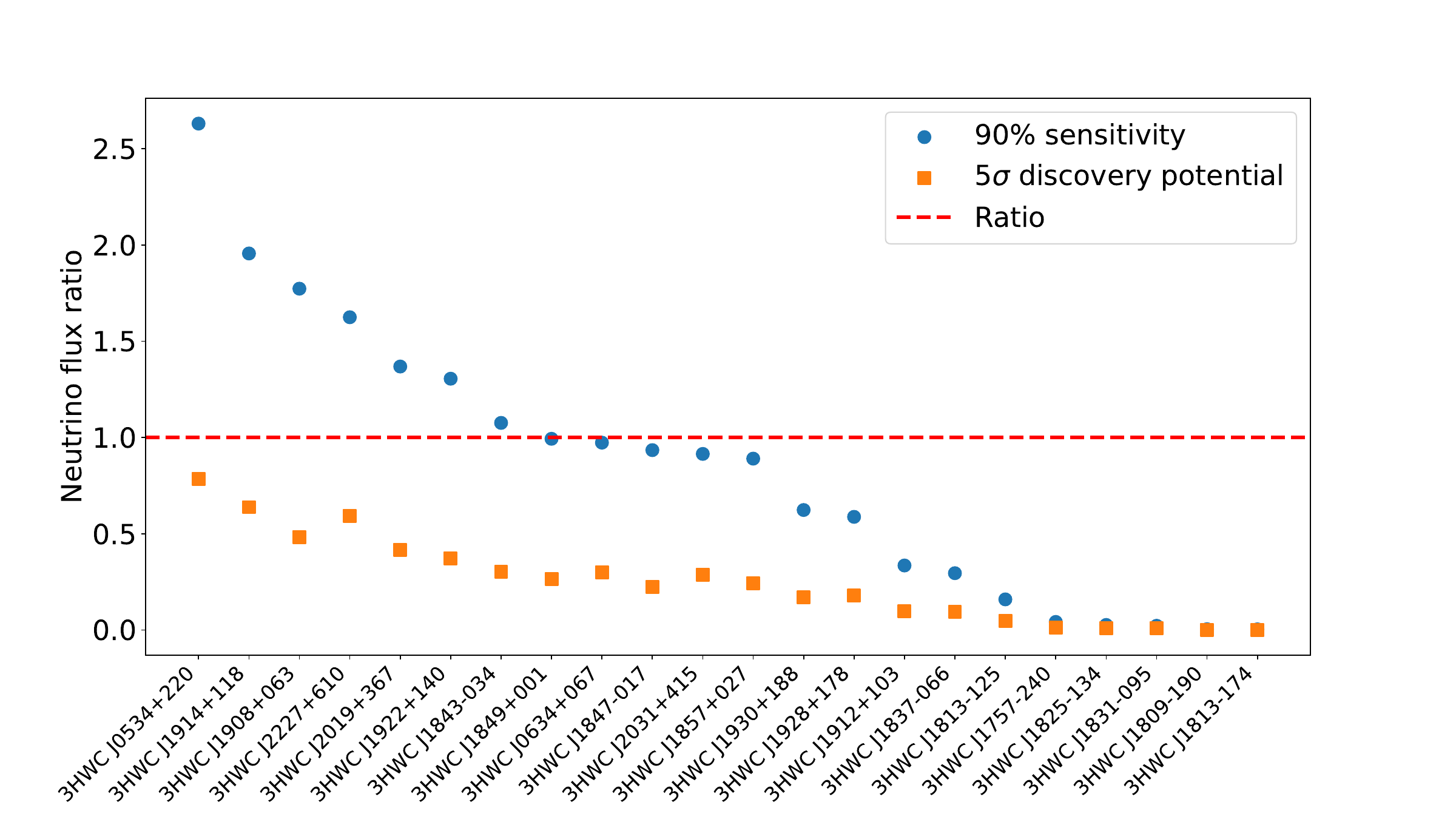}
\caption{The ratio between the predicted neutrino flux and the sensitivity~(blue dot) and 5$\sigma$ discovery potential~(orange rectangle) of the analysis. The sources above the red dashed line are the sources that are likely to be detected by IceCube if they are hadronic sources. The plot is ordered by the sensitivity ratio.}
\label{fig:sensitivity_ratio}
\end{figure}

The background TS distribution is calculated by generating 50,000 background-only trials. For each background trial, the IceCube data are scrambled in RA while keeping the HAWC component of the likelihood unchanged, and the joint analysis is performed. To determine the sensitivity of the joint search for a source with the best-fit parameters, we inject different simulated values of the neutrino flux at the location of the best-fit gamma-ray source with its best-fit extension and spectral shape. The 90\% CL sensitivity is the neutrino flux required for the TS to exceed the median of the background-only TS distribution 90\% of the time. The $5\sigma$ discovery potential is the neutrino flux required for the TS to exceed the $5\sigma$ fluctuation of the background-only TS distribution 50\% of the time. Figure \ref{fig:sensitivity_ratio} shows the ratio between the sensitivity and the predicted neutrino flux assuming pp interactions, as well as the ratio between the $5\sigma$ discovery potential and the predicted neutrino flux assuming pp interactions.

%need citation%
% \begin{figure}[!ht]
% \label{sensitivity}
% \centering
% \includegraphics[width=\textwidth]{plot/sens_predict_withfit.png}
% \caption{The 90\% sensitivity and $5\sigma$ discovery potential of the joint analysis. The orange dot represents the $5\sigma$ discovery potential and the blue triangle represents the sensitivity. The red dot represents the neutrino flux predicted from the gamma-ray best fit assuming all the gamma-ray emission originated from hadronic interaction. The size represents the relative extension of the source and the point source is assigned a 0.1 degree relative size.}
% \end{figure}

We derive the pre-trial p-value for each source by using the background TS distribution of each source. The pre-trial p-value is the percentage of background trials having a TS higher than the TS of the actual search. To correct for the look-elsewhere effect, we create a pre-trial p-value distribution by searching for neutrinos from each source using the same background data and picking the smallest p-value from the 22 sources in each background trial. The post-trial p-value is the fraction of p-values in the aforementioned distribution that are smaller than the smallest pre-trial p-value in the data. With this method, the correlations between different sources can be correctly taken into account. 

%need citation%
In addition to searching for individual sources, we perform a binomial test to determine whether a subset of the sources has TS values inconsistent with the background \citep{grb_icecube}. The test statistic of the binomial test, $p_{binomial}$, is the minimum probability, over all $k$ that at least $k$ p-values are at or below the kth smallest p-value observed, $p_k$:
\begin{equation}
p_{binomial}=\min_k{P_k}=\min_k{\sum_{m=k}^N \binom{N}{m} p_k^m(1-p_k)^{N-m}}
\label{bino}
\end{equation}

To account for the look-elsewhere effect, we create a distribution of $p_{binomial}$ by performing the binomial test on scrambled data. We compute the post-trial p-value of the binomial test using the $p_{binomial}$ value for data. To calculate the overall post-trial p-value accounting for the trials from doing two searches, we create another p-value distribution by choosing, for each background trial, the smaller of the post-trial p-values from the individual source search and the binomial test. This method takes into account the correlation between the individual source search and the binomial test. We then compare the smaller of the two p-values from data with this distribution to obtain the overall p-value.%. and the effective trial factor will be less than two. %We compute the overall post-trial p-value by computing the percentage of trials from this p-value distribution having a smaller p-value than p, where p is the smaller p-value between the post-trial~(only account trials from individual source search) p-value of individual source search and the post-trial~(only account trials from the binomial test) p-value of the binomial test.
%need citation liz paper%

\section{Results}
\label{sec:result}
\subsection{HAWC best-fit result}
 First, we present the best-fit results for the gamma-ray emission from the HAWC Pass 5 data alone. Table \ref{HAWC_best_fit} summarizes the best-fit location, extension, and spectral parameters of each source in our list. In addition, we report the statistical and systematic uncertainty of the spectral fit results. Systematic uncertainties of the HAWC detector include PMT efficiency, charge uncertainty, etc., and are described in \citet{crab100tev}. To evaluate their effect, we perform the model selection using samples simulated with different settings and compare the results with the baseline simulation. The errors of each type of systematic uncertainty are added in quadrature to obtain the total systematic uncertainty, which is given in the table \ref{HAWC_best_fit}. We also reported the $2\sigma$ energy range of each source. The upper end of this range is defined such that if the spectrum is cut off above this energy the significance of the source drops by $2\sigma$. The corresponding definition is used for the lower end of the sensitivity range. The $2\sigma$ threshold is chosen to reasonably claim that HAWC data can probe the quoted energy range at a significance of $2\sigma$ or at the $\sim$95\% confidence level.
%%table xxx

\begin{table}[!ht]

\scriptsize 
    \centering
\begin{tabular}
{c  c c c  c c c c c}

    \hline
        %Name & RA [$^\circ$] & Dec [$^\circ$] & Ext [$^\circ$] & $\alpha$ & $\beta$ & $F_{7}$ [$10^{-14}$TeV$^{-1}$cm$^{-2}$s$^{-1}$] & Nearest TeVCat source \\ \hline
            Name & RA  & Dec & Ext & $\alpha$ & $\beta$ & $F_{7}$  & Nearest TeVCat & Energy range\\ 
            &[$^\circ$]& [$^\circ$]&[$^\circ$] & & &[$10^{-14}$TeV$^{-1}$cm$^{-2}$s$^{-1}$]& &[TeV] \\ \hline
        3HWC J0534+220 & 83.64 & 22.01 & PS & $-$2.82$^{+0.01}_{-0.01}$ $^{+0.09 }_{-0.03}$ & 0.12$^{+0.01}_{-0.01}$ $^{+0.0 }_{-0.05}$ & $23.82^{+0.23}_{-0.22}$ $^{+0}_{-2.7}$ & Crab & 0.1 - 106.1 \\ \hline
        3HWC J0634+067 & 98.59 & 6.66 & 0.49 & $-$2.57$^{+0.1}_{-0.1}$ $^{+0.21 }_{-0.0}$ & 0 & $1.35^{+0.43}_{-0.32}$ $^{+0}_{-2.6}$ & HAWC J0635+070 & 1.7-53.1  \\ \hline
        3HWC J1757$-$240 & 269.32 & $-$23.86 & PS & $-$2.58$^{+0.26}_{-0.26}$ $^{+0.97 }_{-0.11}$ & 0 & $1.68^{+1.03}_{-0.66}$ $^{+0}_{-1.5}$ & HESS J1800$-$240B & 16.9 - 59.7  \\ \hline
        3HWC J1809$-$190 & 272.42 & $-$19.34 & 0.23 & $-$2.09$^{+0.15}_{-0.15}$ $^{+0.01 }_{-0.27}$ & 0.23$^{+0.07}_{-0.07}$ $^{+0.0 }_{-0.3}$ & $11.69^{+1.11}_{-1.06}$ $^{+0}_{-3.1}$ & HESS J1809$-$193 & 3.0 - 94.5  \\ \hline
        3HWC J1813$-$125 & 273.4 & $-$12.7 & 0.27 & $-$2.6$^{+0.08}_{-0.08}$ $^{+0.48 }_{-0.0}$ & 0 & $2.87^{+0.47}_{-0.44}$ $^{+1.0}_{-0}$ & HESS J1813-126 & 2.4 - 59.6 \\ \hline
        3HWC J1813$-$174 & 273.4 & $-$17.74 & 0.26 & $-$1.91$^{+0.16}_{-0.15}$ $^{+0.54 }_{-0.53}$ & 0.35$^{+0.08}_{-0.08}$ $^{+0.31 }_{-0.31}$ & $13.47^{+1.22}_{-1.07}$ $^{+2.5}_{-5.6}$ & 2HWC J1814$-$173& 3.4 - 106.0   \\ \hline
        3HWC J1825$-$134 & 276.49 & $-$13.63 & 0.46 & $-$2.45$^{+0.02}_{-0.02}$ $^{+0.1 }_{-0.02}$ & 0.13$^{+0.02}_{-0.02}$ $^{+0.0 }_{-0.14}$ & $41.02^{+1.12}_{-1.12}$ $^{+0}_{-8.9}$ & 2HWC J1825$-$134& 0.5 - 106.0  \\ \hline
        3HWC J1831$-$095 & 277.85 & $-$9.85 & 1.14 & $-$2.59$^{+0.03}_{-0.04}$ $^{+0.22 }_{-0.0}$ & 0.14$^{+0.02}_{-0.03}$ $^{+0.0 }_{-0.18}$ & $37.37^{+2.08}_{-1.9}$ $^{+0}_{-8.4}$ & HESS J1831$-$098& 0.4 - 118.7   \\ \hline
        3HWC J1837$-$066 & 279.38 & $-$6.83 & 0.65 & $-$2.69$^{+0.02}_{-0.02}$ $^{+0.14 }_{-0.0}$ & 0.1$^{+0.01}_{-0.01}$ $^{+0.0 }_{-0.08}$ & $34.52^{+0.89}_{-0.88}$ $^{+0}_{-4.8}$ & 2HWC J1837$-$065& 0.2 - 133.2  \\ \hline
        3HWC J1843$-$034 & 280.95 & $-$3.39 & 0.6 & $-$2.51$^{+0.03}_{-0.03}$ $^{+0.2 }_{-0.0}$ & 0.13$^{+0.02}_{-0.02}$ $^{+0.0 }_{-0.21}$ & $21.51^{+0.79}_{-0.74}$ $^{+0}_{-8.4}$ & 2HWC J1844$-$032& 0.3 - 75.0   \\ \hline
        3HWC J1847$-$017 & 282.07 & $-$1.78 & 0.75 & $-$2.66$^{+0.03}_{-0.03}$ $^{+0.28 }_{-0.0}$ & 0.09$^{+0.02}_{-0.02}$ $^{+0.0 }_{-0.18}$ & $17.03^{+0.81}_{-0.76}$ $^{+0}_{-2.7}$ & HESS J1848$-$018 & 0.3 - 84.0  \\ \hline
        3HWC J1849+001 & 282.31 & 0.02 & 0.68 & $-$2.42$^{+0.03}_{-0.04}$ $^{+0.22 }_{-0.0}$ & 0.13$^{+0.02}_{-0.02}$ $^{+0.0 }_{-0.24}$ & $15.38^{+0.69}_{-0.7}$ $^{+0}_{-3.9}$ & IGR J18490$-$0000 & 0.4 - 149.8  \\ \hline
        3HWC J1857+027 & 284.35 & 2.82 & 0.6 & $-$2.73$^{+0.02}_{-0.02}$ $^{+0.28 }_{-0.0}$ & 0.13$^{+0.02}_{-0.02}$ $^{+0.0 }_{-0.18}$ & $16.3^{+0.59}_{-0.58}$ $^{+0}_{-3.3}$ & HESS J1857+026 & 0.3 - 133.2  \\ \hline
        3HWC J1908+063 & 287.02 & 6.35 & 0.57 & $-$2.44$^{+0.02}_{-0.02}$ $^{+0.09 }_{-0.01}$ & 0.09$^{+0.01}_{-0.01}$ $^{+0.0 }_{-0.07}$ & $20.02^{+0.52}_{-0.5}$ $^{+0}_{-2.6}$ & MGRO J1908+06 & 0.4 - 168.0  \\ \hline
        3HWC J1912+103 & 288.1 & 10.32 & 0.52 & $-$2.83$^{+0.05}_{-0.05}$ $^{+0.32 }_{-0.0}$ & 0.14$^{+0.04}_{-0.04}$ $^{+0.0 }_{-0.16}$ & $5.79^{+0.47}_{-0.44}$ $^{+2.2}_{-5.6}$ & HESS J1912+101 & 0.4 - 94.4  \\ \hline
        3HWC J1914+118 & 288.71 & 11.79 & 0.71 & $-$2.69$^{+0.04}_{-0.04}$ $^{+0.39 }_{-0.0}$ & 0 & $4.09^{+0.59}_{-0.5}$ $^{+0}_{-2.6}$ & 2HWC J1914+117* & 0.3 - 106.0  \\ \hline
        3HWC J1922+140 & 290.74 & 14.06 & 0.13 & $-$2.77$^{+0.06}_{-0.05}$ $^{+0.12 }_{-0.0}$ & 0 & $1.37^{+0.17}_{-0.12}$ $^{+0.054}_{-0.071}$ & W 51 & 0.2 - 47.3  \\ \hline
        3HWC J1928+178 & 292.13 & 17.81 & 0.83 & $-$2.52$^{+0.04}_{-0.04}$ $^{+0.18 }_{-0.0}$ & 0.11$^{+0.02}_{-0.03}$ $^{+0.0 }_{-0.14}$ & $9.4^{+0.58}_{-0.54}$ $^{+0}_{-1.7}$ & 2HWC J1928+177 & 0.6 - 106.0  \\ \hline
        3HWC J1930+188 & 292.56 & 18.81 & 0.76 & $-$2.51$^{+0.04}_{-0.04}$ $^{+0.12 }_{-0.0}$ & 0.14$^{+0.03}_{-0.03}$ $^{+0.0 }_{-0.12}$ & $9.0^{+0.49}_{-0.48}$ $^{+0}_{-2.2}$ & SNR G054.1+00.3 & 0.7 - 105.9  \\ \hline
        3HWC J2019+367 & 304.9 & 36.77 & 0.32 & $-$2.04$^{+0.05}_{-0.05}$ $^{+0.02 }_{-0.13}$ & 0.31$^{+0.03}_{-0.03}$ $^{+0.0 }_{-0.22}$ & $11.7^{+0.4}_{-0.38}$ $^{+0}_{-2.3}$ & VER J2019+368 & 0.9 - 106.0  \\ \hline
        3HWC J2031+415 & 308.01 & 41.49 & 0.51 & $-$2.52$^{+0.04}_{-0.04}$ $^{+0.26 }_{-0.0}$ & 0.19$^{+0.03}_{-0.03}$ $^{+0.0 }_{-0.28}$ & $12.06^{+0.59}_{-0.56}$ $^{+0}_{-3.7}$ & TeV J2032+4130 & 0.5 - 105.9   \\ \hline
        3HWC J2227+610 & 336.82 & 60.94 & PS & $-$2.42$^{+0.2}_{-0.21}$ $^{+0.78 }_{-0.27}$ & 0 & $1.53^{+0.68}_{-0.47}$ $^{+0.5}_{-1.5}$ & Boomerang & 16.7 - 33.5   \\ \hline

    \end{tabular}
\caption{The best-fit result from the HAWC data. The columns from left to right are 3HWC source name, Right Ascension, Declination, Gaussian extension~(PS indicates a point source), $\alpha$ and $\beta$ of the log-parabola spectrum~($\beta = 0$ for power law spectrum), the differential gamma-ray flux at 7 TeV, , the nearest TeVCat source listed in 3HWC (\citealp{3hwc,TeVcat}) and the $2\sigma$ energy range. The first uncertainty is statistical and the second is systematics.}   
\label{HAWC_best_fit}
\end{table}
\subsection{Joint-fit result}
We performed the joint fit using the best-fit model of HAWC and added a neutrino source as described in Section \ref{joint-fit}. We found no significant neutrino emission. The smallest pre-trial p-value from the individual source search is 0.07 for 3HWC J2019+367, which corresponds to a post-trial p-value of 0.21 when accounting for the trials in the individual source search. We performed the binomial test and found no significant neutrino emission from any subset of the 22 sources. The smallest pre-trial p-value of the binomial test is 0.09 at $k$=6 (3HWC J2019+367, 3HWC J2031+415, 3HWC J1912+103, 3HWC J1908+063, 3HWC J1857+027, 3HWC J1928+178) and corresponds to a post-trial p-value of 0.34 when only accounting for trials in the binomial test. Accounting for both the trials from the individual source search and the binomial test, the final post-trial p-value is 0.37. We also computed the 90\% neutrino energy range for each source using the best-fit morphology and spectrum. The 90\% neutrino energy range is the central 90\% range of true neutrino energy when injecting signal neutrinos with the source's best-fit gamma-ray morphology and spectrum. IceCube applies a higher energy threshold for neutrino events from the southern sky due to the containment of atmospheric muon, resulting in a higher energy range for southern sources.

\begin{figure}[h!]

\centering
\includegraphics[width=\textwidth]{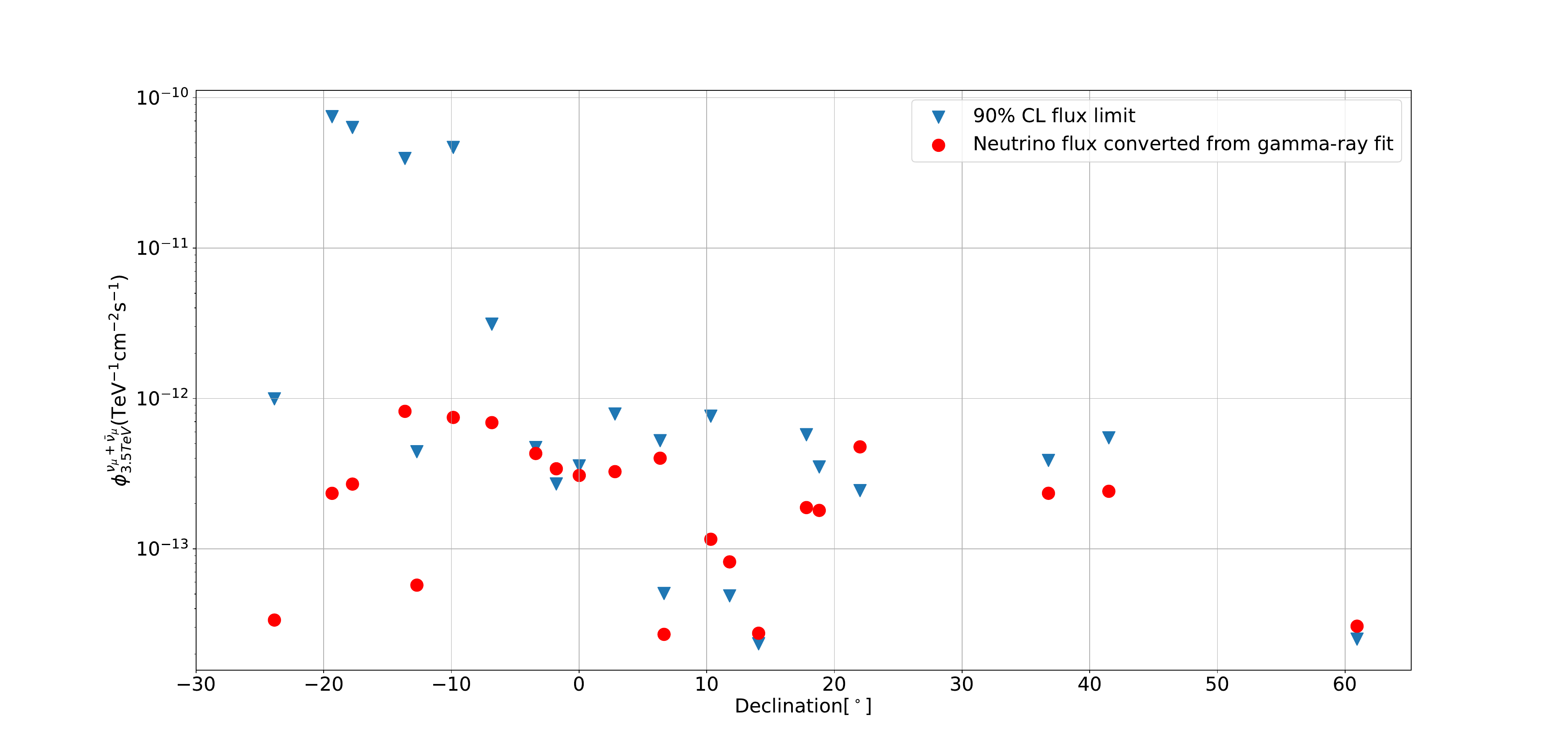}
\caption{Neutrino 90\% CL flux limit from the individual source search. The blue triangles represent the flux limit and the red dots represent the neutrino flux predicted from the gamma-ray best fit assuming all the gamma-ray emission originated from hadronic interaction. Sources that have a predicted neutrino flux higher than the flux limit are the sources for which we can place a hadronic fraction constraint.}
\label{flux_limit}
\end{figure}

Based on the null result of the joint search, we also calculated the neutrino flux limits at 90\% CL for the 22 sources (see table \ref{flux_limit_table}). We account for IceCube systematics by calculating the sensitivity using different Monte Carlo systematic sets and adding the difference to the baseline in quadrature. Appendix \ref{sys_table_section} details the effect of IceCube systematics on the flux limit for each source. Table \ref{flux_limit_table} shows the neutrino flux limit for each source. Figure \ref{flux_limit} shows, by declination, the neutrino flux limit and the neutrino flux predicted from the best gamma-ray fit assuming all the gamma-ray emission is from hadronic interaction.

\begin{table}[!ht]
    \centering
   \footnotesize
\begin{tabular}
{c  c c c c  c c c}
    \hline

      Name &  RA  &  Dec &  Extension & p-value &  Neutrino 90\% CL flux limit  & Predicted neutrino flux & Energy range\\ 
      & {\footnotesize[$^\circ$]}&{\footnotesize [$^\circ$]} &{\footnotesize[$^\circ$}]& &{\footnotesize[TeV$^{-1}$cm$^{-2}$s$^{-1}$]} & {\footnotesize[TeV$^{-1}$cm$^{-2}$s$^{-1}$]} & {\footnotesize{[TeV]}} \\ \hline

3HWC J0534+220 & 83.64 & 22.01 & PS &0.36& $2.44		\times  10^{-13} $ & $4.74\times  10^{-13} $ & 0.3 - 21.6\\ \hline
3HWC J0634+067 & 98.59 & 6.66 & PS &0.19& $5.04	\times  10^{-14}$ & $2.74\times  10^{-14} $ & 0.4 - 143.0 \\ \hline
3HWC J1757$-$240 & 269.32 & $-$23.86 & 0.10 &1& $9.94	\times  10^{-13} $ & $3.35\times  10^{-14} $ & 98.7 - 6870.0 \\ \hline
3HWC J1809$-$190 & 272.42 & $-$19.34 & 0.23 &1& $7.47  \times  10^{-11}$ & $2.32\times  10^{-13} $ & 38.7 - 468.3\\ \hline
3HWC J1813$-$125 & 273.40 & $-$12.70 & 0.27 &1& $4.42	\times  10^{-13} $ & $5.55\times  10^{-14} $ & 41.3 - 3897.3 \\ \hline
3HWC J1813$-$174 & 273.40 & $-$17.74 & 0.26 &0.7& $6.33	\times  10^{-11}$ & $2.72\times  10^{-13} $ & 9.60 - 263.5\\ \hline
3HWC J1825$-$134 & 276.49 & $-$13.63 & 0.46&1 & $3.94	\times  10^{-11} $ & $8.00\times  10^{-13} $ & 26.2 - 519.7 \\ \hline
3HWC J1831$-$095 & 277.85 & $-$9.85 & 1.14 &1& $4.71	\times  10^{-11} $ & $7.39\times  10^{-13} $ & 3.1 - 229.4\\ \hline
3HWC J1837$-$066 & 279.38 & $-$6.83 & 0.65 &1& $3.12	\times  10^{-12} $ & $6.91\times  10^{-13} $ & 0.7 - 94.1\\ \hline
3HWC J1843$-$034 & 280.95 & $-$3.39 & 0.60 &1& $4.71	\times  10^{-13} $ & $4.28\times  10^{-13} $ & 0.7 - 57.0 \\ \hline
3HWC J1847$-$017 & 282.07 & $-$1.78 & 0.75 &1& $2.70	\times  10^{-13} $ & $3.45\times  10^{-13} $ & 0.5 - 49.1\\ \hline
3HWC J1849+001 & 282.31 & 0.02 & 0.68&1 & $3.56	\times  10^{-13} $ & $3.08\times  10^{-13} $ & 0.7 - 56.9 \\ \hline
3HWC J1857+027 & 284.35 & 2.82 & 0.60 &0.13& $7.87	\times  10^{-13} $ & $3.26\times  10^{-13} $ & 0.4 - 29.2 \\ \hline
3HWC J1908+063 & 287.02 & 6.35 & 0.57 &0.1& $5.24	\times  10^{-13} $ & $4.00\times  10^{-13} $ & 0.6 - 65.1 \\ \hline
3HWC J1912+103 & 288.10 & 10.32 & 0.52 &0.08& $7.61	\times  10^{-13} $ & $1.11\times  10^{-13} $ & 0.4 - 24.1\\ \hline
3HWC J1914+118 & 288.71 & 11.79 & 0.71 &1& $4.87	\times  10^{-14} $ & $7.93\times  10^{-14} $ & 0.3 - 73.9\\ \hline
3HWC J1922+140 & 290.74 & 14.06 & 0.13 &1& $2.33	\times  10^{-14}$ & $2.68\times  10^{-14} $ & 0.3 - 51.8\\ \hline
3HWC J1928+178 & 292.13 & 17.81 & 0.83 &0.14& $5.73	\times  10^{-13} $ & $1.85\times 10^{-13} $ & 0.5 - 42.7\\ \hline
3HWC J1930+188 & 292.56 & 18.81 & 0.76 &0.1& $3.50	\times  10^{-13} $ & $1.78\times  10^{-13} $ & 0.5 - 36.6 \\ \hline
3HWC J2019+367 & 304.90 & 36.77 & 0.32 &0.07& $3.87	\times  10^{-13} $ & $2.34\times  10^{-13} $ & 1.1 - 34.2\\ \hline
3HWC J2031+415 & 308.01 & 41.49 & 0.51 &0.08& $5.47	\times  10^{-13} $ & $2.40\times  10^{-13} $ & 0.5 - 25.8 \\ \hline
3HWC J2227+610 & 336.82 & 60.94 & PS &1& $2.51	\times  10^{-14}$ & $3.11\times  10^{-14} $ & 0.4 - 78.4\\ \hline

    \end{tabular}
\caption{Neutrino 90\% CL flux limit at 3.5 TeV from the individual source search. The neutrino flux limit is calculated by injecting a flux of neutrino according to the gamma-ray best-fit model's neutrino prediction. The predicted neutrino flux is calculated from the gamma-ray fit result assuming the emission is purely hadronic. The IceCube 90\% energy range is the central 90\% energy range of the signal neutrinos. The smallest pre-trial p-value is 0.07 and it corresponds to 0.21 post-trial p-value after accounting for trials in the individual source search.}  
\label{flux_limit_table}
\end{table}

Five sources have a lower neutrino 90\% CL flux limit than the predicted neutrino flux. In these cases, we conclude that the gamma-ray emission detected in the HAWC energy range cannot completely originate from hadronic interactions. This assumes that the HAWC observed spectrum extends to the IceCube energy range and it doesn't account for HAWC spectral uncertainty. If we further assume that the leptonic component shares a similar spectral shape in the TeV energy range, we can constrain the percentage of the gamma-ray flux originating from hadronic interactions. Among these sources, 3HWC J2227+610 was studied by the HAWC collaboration and could be associated with supernova remnant G106.3+2.7. HAWC found a hint for PeV proton acceleration and suggested that 3HWC J2227+610 could be a Galactic PeVatron~\citep{j2227}. The Crab Nebula (3HWC J0534+220) is the brightest source in the TeV sky and is the most strongly constrained. However, the Crab Nebula is widely believed to be a leptonic source. A more sensitive neutrino instrument is required to put a relevant constraint, in the $\sim $1\% range, on the hadronic fraction. 3HWC J1922+140 is close to the W 51 region which contains the star-forming region W51B, the supernova remnant W51C, and a potential PWN, CXOU J192318.5+140305 \citep{HESS_GP}. 3HWC J1922+140 could be a composite of different astrophysical origins. Table \ref{fraction_table} shows the constraints on the hadronic fraction, which range from 51\% to 85\%.
%% insert table and figure
\begin{table}[ht!]

\begin{tabular}{cccccc}
\hline
Name         & $\alpha$ & $\beta$ & Neutrino 90\% CL flux limit [TeV$^{-1}$cm$^{-2}$s$^{-1}$]
 & p-value & Hadronic fraction limit \\ \hline
%&&&&&\\\hline
3HWC J1847-017 & $-$2.66 & 0.09 & $2.70\times 10^{-13}$               & 1       & 0.79        \\ \hline
3HWC J1914+118 & $-$2.69 & 0    & $4.87\times 10^{-14}$               & 1       & 0.59        \\ \hline
3HWC J1922+140 & $-$2.77 & 0    & $2.33\times 10^{-14}$               & 1       & 0.85        \\ \hline
3HWC J0534+220 & $-$2.82 & 0.12 & $2.44\times 10^{-13}$               & 0.36 & 0.51        \\ \hline
3HWC J2227+610 & $-$2.42 & 0    & $2.51\times 10^{-14}$               & 1       & 0.82        \\ \hline
\end{tabular}
            \caption{Table showing the sources that have a neutrino 90\% CL flux limit lower than the predicted neutrino flux at 3.5 TeV when assuming the source is hadronic. The p-value is the pre-trial p-value from the individual source search and a p-value of 1 corresponds to a TS of 0 in the search. The hadronic fraction limit is the ratio between the flux limit and the predicted neutrino flux at 3.5 TeV, representing the maximum hadronic fraction assuming the leptonic emission shares a similar spectral shape.}  \label{fraction_table}
\end{table}
\section{Summary and discussion}
\label{sec:summary}
In this paper, we present a new method for performing multi-messenger spectral fits with the 3ML software framework \citep{3ML} using HAWC and IceCube data. This is motivated by the predicted neutrino emission by Galactic cosmic-ray accelerators with gamma-ray emission. The new joint-fitting method with gamma-ray data and neutrino data brings improvement in search sensitivity due to more detailed modeling of the neutrino emission under the assumption that the gamma rays and neutrinos originated from the same population of hadrons. In addition, we could provide a direct constraint on the hadronic ratio in the case of null detection using the emission model derived from the gamma-ray data if the assumption is true. We stress the caveat that a mis-modeling of the emission region, for example, due to gamma rays and neutrinos being produced in different regions of a source, could weaken the constraints presented in this paper.

We used the HAWC data to obtain the best-fit model for the gamma-ray emission. We performed a joint fit using HAWC and IceCube data to search for neutrino emission from potential Galactic cosmic-ray accelerators. No significant neutrino emission was observed. The most significant source from the joint fit is 3HWC J2019+367 with the lowest pre-trial p-value of 0.07, corresponding to a post-trial p-value of 0.21. A binomial test for an excess distributed over several sources was performed. The binomial test gave a pre-trial p-value of 0.09, obtained for the six most significant sources and corresponds to a post-trial p-value of 0.34. After trial correction for the two searches, the overall post-trial p-value is 0.37. We calculated upper limits at 90\% confidence level on the neutrino flux, assuming the spectral form corresponding to the best fit for gamma rays. We found that five sources have a neutrino flux limit that is lower than the predicted neutrino flux of a purely hadronic source. We conclude that these five sources cannot be purely hadronic and constrain the hadronic fraction of these sources. 

 We found that for some of the sources, the HAWC data do not show a 2$\sigma$
significance for emission above 100 TeV, meaning that the analysis would not warrant the conclusion that those sources are PeVatrons even if neutrinos are detected from the source. In addition, the neutrino 90\% energy range for some of the sources does not exceed 50 TeV, meaning the neutrino data would likely not be grounds to establish PeV cosmic-ray acceleration in the case of detection. A more sensitive experiment such as LHAASO could potentially determine the limit on the maximum energy of the source, and neutrino detection could act as evidence of hadronic interactions. In addition, it could provide more PeVatron candidates with better gamma-ray spectrum and morphology measurement. The recently released 1LHAASO catalog contains 43 sources with gamma-ray emission above 100 TeV \citep{LHAASO_first_catalog} and a joint search with neutrino data will provide better sensitivity and constraints on their hadronic emission. Next-generation neutrino detectors can improve our understanding of the physical processes of these TeV gamma-ray emitters. The proposed IceCube-Gen2 detector \citep{ICgen2}, with an increased effective area compared to IceCube, is expected to become up to five times more sensitive to galactic neutrino sources. Several neutrino experiments, such as the proposed Pacific Ocean Neutrino Experiment~(P-ONE) \citep{pone}, Tropical Deep-sea Neutrino Telescope~(TRIDENT) \citep{TRIDENT}, KM3NeT~(currently under construction) \citep{km3net} and Baikal-GVD~(operating and undergoing upgrade) \citep{Baikal}, are located in the northern hemisphere and have the potential to detect galactic neutrino sources near the galactic center. The proposed SWGO experiment, which is a ground-based gamma-ray water Cherenkov observatory in the southern hemisphere, would be able to discover more TeV gamma rays sources in the southern sky and candidates of Galactic PeVatrons \citep{swgo}.

\begin{acknowledgments}
HAWC: We acknowledge the support from: the US National Science Foundation (NSF); the US Department of Energy Office of High-Energy Physics; the Laboratory Directed Research and Development (LDRD) program of Los Alamos National Laboratory; Consejo Nacional de Ciencia y Tecnolog\'{i}a (CONACyT), M\'{e}xico, grants 271051, 232656, 260378, 179588, 254964, 258865, 243290, 132197, A1-S-46288, A1-S-22784, CF-2023-I-645, c\'{a}tedras 873, 1563, 341, 323, Red HAWC, M\'{e}xico; DGAPA-UNAM grants IG101323, IN111716-3, IN111419, IA102019, IN106521, IN114924, IN110521 , IN102223; VIEP-BUAP; PIFI 2012, 2013, PROFOCIE 2014, 2015; the University of Wisconsin Alumni Research Foundation; the Institute of Geophysics, Planetary Physics, and Signatures at Los Alamos National Laboratory; Polish Science Centre grant, DEC-2017/27/B/ST9/02272; Coordinaci\'{o}n de la Investigaci\'{o}n Cient\'{i}fica de la Universidad Michoacana; Royal Society - Newton Advanced Fellowship 180385; Generalitat Valenciana, grant CIDEGENT/2018/034; The Program Management Unit for Human Resources \& Institutional Development, Research and Innovation, NXPO (grant number B16F630069); Coordinaci\'{o}n General Acad\'{e}mica e Innovaci\'{o}n (CGAI-UdeG), PRODEP-SEP UDG-CA-499; Institute of Cosmic Ray Research (ICRR), University of Tokyo; National Research Foundation of Korea (RS-2023-00280210). H.F. acknowledges support by NASA under award number 80GSFC21M0002. We also acknowledge the significant contributions over many years of Stefan Westerhoff, Gaurang Yodh and Arnulfo Zepeda Dom\'inguez, all deceased members of the HAWC collaboration. Thanks to Scott Delay, Luciano D\'{i}az and Eduardo Murrieta for technical support.

IceCube: The IceCube collaboration acknowledges the significant contributions to this manuscript from Kwok Lung Fan. The authors gratefully acknowledge the support from the following agencies and institutions: USA – U.S. National Science Foundation-Office of Polar Programs, U.S. National Science Foundation-Physics Division, U.S. National Science Foundation-EPSCoR, U.S. National Science Foundation-Office of Advanced Cyberinfrastructure, Wisconsin Alumni Research Foundation, Center for High Throughput Computing (CHTC) at the University of Wisconsin–Madison, Open Science Grid (OSG), Partnership to Advance Throughput Computing (PATh), Advanced Cyberinfrastructure Coordination Ecosystem: Services \& Support (ACCESS), Frontera computing project at the Texas Advanced Computing Center, U.S. Department of Energy-National Energy Research Scientific Computing Center, Particle astrophysics research computing center at the University of Maryland, Institute for Cyber-Enabled Research at Michigan State University, Astroparticle physics computational facility at Marquette University, NVIDIA Corporation, and Google Cloud Platform; Belgium – Funds for Scientific Research (FRS-FNRS and FWO), FWO Odysseus and Big Science programmes, and Belgian Federal Science Policy Office (Belspo); Germany – Bundesministerium für Bildung und Forschung (BMBF), Deutsche Forschungsgemeinschaft (DFG), Helmholtz Alliance for Astroparticle Physics (HAP), Initiative and Networking Fund of the Helmholtz Association, Deutsches Elektronen Synchrotron (DESY), and High Performance Computing cluster of the RWTH Aachen; Sweden – Swedish Research Council, Swedish Polar Research Secretariat, Swedish National Infrastructure for Computing (SNIC), and Knut and Alice Wallenberg Foundation; European Union – EGI Advanced Computing for research; Australia – Australian Research Council; Canada – Natural Sciences and Engineering Research Council of Canada, Calcul Québec, Compute Ontario, Canada Foundation for Innovation, WestGrid, and Digital Research Alliance of Canada; Denmark – Villum Fonden, Carlsberg Foundation, and European Commission; New Zealand – Marsden Fund; Japan – Japan Society for Promotion of Science (JSPS) and Institute for Global Prominent Research (IGPR) of Chiba University; Korea – National Research Foundation of Korea (NRF); Switzerland – Swiss National Science Foundation (SNSF).
\end{acknowledgments}


\vspace{5mm}

%% citation or url to the code inside~()s when available.

% \software{astropy \citep{2013A&A...558A..33A,2018AJ....156..123A},  
%           Cloudy \citep{2013RMxAA..49..137F}, 
%           Source Extractor \citep{1996A&AS..117..393B}
%           }


%% Each Appendix~(indicated with \section) will be lettered A, B, C, etc.
%% command and will number appendix equations~(A1),~(A2), etc. The

\appendix

\section{Likelihood formalism}
\label{likelihood_appendix}
HAWC performs a binned likelihood analysis for studying gamma-ray sources. In this analysis, we use the energy binning, the fraction of PMTs hit~($\it fhit$) binning, and also Healpix Nside=1024 spatial binning. We calculate the expected background for each bin by applying direct integration using the zenith distribution of all events passing the gamma/hadron cuts. The technique of direct integration is described in previous HAWC publications \citep{crab2019}. The expected excess in every bin is calculated based on the physics model and the detector response of HAWC. The total log-likelihood is the sum of the Poisson log-likelihood of each bin and pixel.
\begin{equation}
    \ln \mathcal{L}_{\it HAWC} = \sum^{N_{\it bin}}_{j} \sum^{N_{\it pixels}}_{i} \ln \frac{(b_{ij} + e_{ij} f)^{d_{ij}}e^{-(b_{ij}+e_{ij}f)}}{d_{ij}!}
\end{equation}
where $b_{ij}$ is the expected background at bin $j$ and pixel $i$, $e_{ij}$ is the expected excess per unit flux given the model at bin $j$ and pixel $i$, $f$ is the flux and $d_{ij}$ is the actual count in bin $j$ and pixel $i$. Index $j$ loops over the energy and $fhit$ bin and index $i$ loops over the number of pixels in the region of interest.

For IceCube, we perform an unbinned likelihood analysis. The total likelihood is given by
\begin{equation}
\mathcal{L}_{\it IceCube} = \prod_i^N L_i \left(\vec{\theta},\vec{D_i}\right) = \prod_i^N\left(\frac{n_s}{N}S~(\vec{\theta},\vec{D_i}) + \frac{N-n_s}{N}B(\vec{D_i})\right),
\end{equation}
where $\vec{\theta}$ represents the properties of the source hypothesis like location, morphology, and spectrum. $L_i$ is the likelihood of the event $i$. The parameter vector $\vec{D_i}$ represents the reconstruction information of the event $i$, including reconstructed energy, direction, and estimated angular error. $N$ is the total number of events and $n_s$ is the number of signal neutrinos. $S$ and $B$ are the signal probability density function (PDF) and background probability density function, each consisting of a product of a spatial likelihood and an energy likelihood. A detailed description of the IceCube likelihood for extended sources can be found in \cite{Fan_chang_icecube}. Since the background likelihood only depends on the reconstructed information of the event and does not depend on the source hypothesis, it is a constant during the fitting process. Therefore, we can divide the likelihood by the likelihood of the background-only hypothesis to get the likelihood ratio which is still a valid likelihood. We rewrite the IceCube log-likelihood to be 
\begin{equation} 
\ln \mathcal{L}_{\it IceCube}= \sum_i^N \ln\left( \frac{n_s}{N}\left(\frac{S(\vec{D_i},\vec{\theta})}{B(\vec{D_i})}-1\right)+1\right).
\end{equation}

Since HAWC and IceCube measurements are independent, the total log-likelihood for the joint fit is the addition of the HAWC log-likelihood and IceCube log-likelihood.
\begin{equation} 
\ln \mathcal{L}_{HAWC} + \ln \mathcal{L}_{\it IceCube}= \sum^{N_{\it bin}}_{j} \sum^{N_{\it pixels}}_{i} \ln \frac{(b_{ij} + e_{ij} f)^{d_{ij}}e^{-(b_{ij}+e_{ij}f)}}{d_{ij}!}+\sum_i^N \ln\left( \frac{n_s}{N}\left(\frac{S(\vec{D_i},\vec{\theta}}{B(\vec{D_i})}-1\right)+1\right).
\end{equation}
The combined log-likelihood is then maximized over the free parameters in the model.
% \section{Pre-trial p-values of the joint search}
% \label{pre-trial-p-value}
% Table \ref{pre-trial-p-value-tab} shows the pre-trial p-value of each source.
% \begin{table}[!ht]
    
%     \centering
%     \begin{tabular}{ c c c}
    
%     \hline
%         Name & pre-trial p-value & post-trial p-value \\ \hline
%         3HWC J0534+220 & 0.36 & - \\ \hline
%         3HWC J0634+067 & 0.19 & - \\ \hline
%         3HWC J1757-240 & 1 & - \\ \hline
%         3HWC J1809-190 & 1 & - \\ \hline
%         3HWC J1813-125 & 1 & - \\ \hline
%         3HWC J1813-174 & 0.70 & - \\ \hline
%         3HWC J1825-134 & 1 & - \\ \hline
%         3HWC J1831-095 & 1 & - \\ \hline
%         3HWC J1837-066 & 1 & - \\ \hline
%         3HWC J1843-034 & 1 & - \\ \hline
%         3HWC J1847-017 & 1 & - \\ \hline
%         3HWC J1849+001 & 1 & - \\ \hline
%         3HWC J1857+027 & 0.13 & - \\ \hline
%         3HWC J1908+063 & 0.10 & - \\ \hline
%         3HWC J1912+103 & 0.08 & - \\ \hline
%         3HWC J1914+118 & 1 & - \\ \hline
%         3HWC J1922+140 & 1 & - \\ \hline
%         3HWC J1928+178 & 0.14 & - \\ \hline
%         3HWC J1930+188 & 1 & - \\ \hline
%         3HWC J2019+367 & 0.07 & 0.21 \\ \hline
%         3HWC J2031+415 & 0.08 & - \\ \hline
%         3HWC J2227+610 & 1 & - \\ \hline
%     \end{tabular}
%     \caption{The pre-trial p-value and the post-trial p-value of the joint fit. A p-value of 1 corresponds to a TS equal to 0.}
%     \label{pre-trial-p-value-tab}
% \end{table}
\section{IceCube Systematics}
\label{sys_table_section}
In this analysis, we consider the effect of uncertainties on the scattering and absorption of light in the ice, uncertainty in DOM efficiency, and uncertainty in the modeling of ice properties. We consider $\pm5\%$ uncertainty in absorption and scattering, $\pm10\%$ in DOM efficiency, and uncertainty in hole ice modeling parameters as described in \citet{holeice_icrc}. The method of handling the systematic uncertainties is similar to the one previously used in IceCube's measurement of the diffuse neutrino flux  \citep{iceCube_diffuse}. We use different simulation data sets, each with the value of one parameter changed to account for its systematic uncertainty. The change in sensitivity for the joint analysis as compared to the baseline was calculated for the different sources of systematic uncertainties, and these changes were added in quadrature. Table \ref{sys_table} shows the total systematic uncertainty on the neutrino 90\% CL flux limit from each source. The final neutrino flux limit of each source is increased by the total systematic uncertainty and is shown in table \ref{flux_limit_table}.

\begin{table}[!ht]
    
    \centering
    \begin{tabular}{c c}
    \hline
        Name & Total systematic uncertainties\\ \hline
        3HWC J0534+220 & 18\% \\ \hline
        3HWC J0634+067 & 25\% \\ \hline
        3HWC J1757-240 & 25\% \\ \hline
        3HWC J1809-190 & 39\% \\ \hline
        3HWC J1813-125 & 27\% \\ \hline
        3HWC J1813-174 & 54\% \\ \hline
        3HWC J1825-134 & 30\% \\ \hline
        3HWC J1831-095 & 40\% \\ \hline
        3HWC J1837-066 & 33\% \\ \hline
        3HWC J1843-034 & 19\% \\ \hline
        3HWC J1847-017 & 21\% \\ \hline
        3HWC J1849+001 & 19\% \\ \hline
        3HWC J1857+027 & 18\% \\ \hline
        3HWC J1908+063 & 18\% \\ \hline
        3HWC J1912+103 & 21\% \\ \hline
        3HWC J1914+118 & 20\% \\ \hline
        3HWC J1922+140 & 14\% \\ \hline
        3HWC J1928+178 & 13\% \\ \hline
        3HWC J1930+188 & 23\% \\ \hline
        3HWC J2019+367 & 14\% \\ \hline
        3HWC J2031+415 & 16\% \\ \hline
        3HWC J2227+610 & 30\% \\ \hline
    \end{tabular}
    \caption{The total IceCube systematics for each source. We add 1 to the numbers and then multiply by the neutrino 90\% CL flux limit computed from baseline Monte Carlo to obtain the final neutrino flux limit.}
    \label{sys_table}
\end{table}
\clearpage
\bibliography{sample631}{}
\bibliographystyle{aasjournal}



\end{document}